\documentclass{article}

\usepackage[numbers]{natbib}         
\usepackage[colorlinks]{hyperref}    
\usepackage[english]{babel} 
\usepackage{amssymb}
\usepackage{amsmath}
\usepackage{txfonts}
\usepackage{mathdots}
\usepackage[classicReIm]{kpfonts}
\usepackage[dvips]{graphicx} 
\usepackage[a4paper, portrait, margin=1in]{geometry}
\usepackage{tabularx}
\usepackage{longtable}
\usepackage{multirow}
\usepackage{booktabs}
\usepackage[labelsep=period]{caption}
\usepackage{makecell}
\begin{document}
	\setlength{\parindent}{0pt}
	\setlength{\parskip}{1ex}
	
	\textbf{\Large Deformable Image Registration using Unsupervised Deep Learning for CBCT-guided Abdominal Radiotherapy}
	
	\bigbreak

	Huiqiao Xie, Yang Lei, Yabo Fu, Tonghe Wang, Justin Roper, Jeffrey D. Bradley, 
	Pretesh Patel, Tian Liu and Xiaofeng Yang*
	
	Department of Radiation Oncology and Winship Cancer Institute, Emory University, Atlanta, GA

	\bigbreak
	\bigbreak
	\bigbreak

	\textbf{*Corresponding author: }
	
	Xiaofeng Yang, PhD
	
	Department of Radiation Oncology
	
	Emory University School of Medicine
	
	1365 Clifton Road NE
	
	Atlanta, GA 30322
	
	E-mail: xiaofeng.yang@emory.edu

	\bigbreak
	\bigbreak
	\bigbreak
	\bigbreak
	\bigbreak
	\bigbreak

	\textbf{Abstract}

	\textbf CBCTs in image-guided radiotherapy provide crucial anatomy information for patient setup and plan evaluation. Longitudinal CBCT image registration could quantify the inter-fractional anatomic changes, e.g. tumor shrinkage, daily OAR variation throughout the course of treatment. The purpose of this study is to propose an unsupervised deep learning based CBCT-CBCT deformable image registration which enables quantitative anatomic variation analysis. The proposed deformable registration workflow consists of training and inference stages that share the same feed-forward path through a spatial transformation-based network (STN). The STN consists of a global generative adversarial network (GlobalGAN) and a local GAN (LocalGAN) to predict the coarse- and fine-scale motions, respectively. The network was trained by minimizing the image similarity loss and the deformable vector field (DVF) regularization loss without the supervision of ground truth DVFs. During the inference stage, patches of local DVF were predicted by the trained LocalGAN and fused to form a whole-image DVF. The local whole-image DVF was subsequently combined with the GlobalGAN generated DVF to obtain final DVF. The proposed method was evaluated using 100 fractional CBCTs from 20 abdominal cancer patients in the experiments and 105 fractional CBCTs from a cohort of 21 different abdominal cancer patients in a holdout test. Qualitatively, the registration results show great alignment between the deformed CBCT images and the target CBCT image. Quantitatively, the average target registration error (TRE) calculated on the fiducial markers and manually identified landmarks was 1.91±1.11 mm. The average mean absolute error (MAE), normalized cross correlation (NCC) between the deformed CBCT and target CBCT were 33.42±7.48 HU, 0.94±0.04, respectively. In summary, an unsupervised deep learning-based CBCT-CBCT registration method is proposed and its feasibility and performance in fractionated image-guided radiotherapy is investigated. This promising registration method could provide fast and accurate longitudinal CBCT alignment to facilitate inter-fractional anatomic changes analysis and prediction.

	\bigbreak
	\bigbreak
	
	\textbf{keywords:} Deformable Image Registration, CBCT, Radiotherapy, Deep Learning.

	\noindent 
	\section{ INTRODUCTION}
	
	The treatment course of fractionated radiotherapy usually lasts weeks or months, during which the shape, position and size of tumor targets and organs at risk (OARs) may vary to different extents. It is essential to account for these changes for accurate dose delivery \cite{RN2, RN1}. Cone-beam computed tomography (CBCT) is commonly used in image guided radiotherapy for patient setup and treatment evaluation prior to beam delivery. Compared to the treatment planning CT, CBCT images usually have poor image quality and soft tissue contrast due to scattering and beam hardening. Since the CBCT scan usually takes longer time than the planning CT, the CBCT images are also susceptible to motion artifacts. As a result, CBCTs are primarily used to verify the alignment with planning CT. The ability of CBCT to track patients’ anatomic changes throughout the treatment course has been under-explored. CBCT could enable detailed inter-fractional anatomic changes evaluation on treatment day throughout the treatment course. As such, inter-fractional target and OAR variation could potentially be modeled to predict anatomic variation in future treatment or to guide treatment planning, such as margin definition, to provide sufficient target coverage and OAR sparing. It is important to provide a CBCT-CBCT image registration tool to quantify the inter-fractional anatomic changes to facilitate such modeling.
	
	Conventional intensity-based deformable image registrations (DIRs), such as optical flow \cite{RN3}, demons \cite{RN4} and viscous fluid model \cite{RN5}, are iterative and generally very slow especially for large datasets. Over-smoothed deformation vector fields (DVFs) are usually produced because these methods apply spatial filters repeatedly throughout the iteration process \cite{RN7, RN6}. The large appearance variances and low image contrast of CBCT pose additional challenges for accurate registration. Landmarks identified either automatically or manually have been used to guide planning CT-CBCT and MRI-CBCT registration \cite{RN9, RN8}. However, the landmark identification process can be challenging and laborious in the presence of severe artifacts, which in turn degrade the landmark-guided DIR \cite{RN10}. Partly due to the poor CBCT image quality, very few papers have been published on CBCT-CBCT image registration \cite{RN13, RN11, RN3, RN12}. Noe et. al. \cite{RN3} tried to accelerate an optical flow method \cite{RN14} on a graphics programing unit (GPU) and the test on CBCT-CBCT registration achieved run time of 64 s for image size of 512 × 512 × 55 and target registration errors (TREs) of 1.8±1.0 mm after rigid registration and 1.6±0.8 mm after deformable registration. Nithiananthan et. al. \cite{RN11} studied the accuracy and convergence of multiscale Demons image registration and they achieved run time of 270 s (image size was not specified) and TRE of 1.6±0.9 mm on the CBCT images of ten head and neck cancer patients. Zachiu et. al. \cite{RN12} implemented and evaluated an Evolution method \cite{RN15}, which estimates the deformation between two images by matching similar contrast patterns instead of pixel intensities, for both CT-CBCT and CBCT-CBCT image registration. They achieved run time of approximately 60 s for registration of images of size 256 × 256 × 256. Jiang et. al. \cite{RN13} proposed an multi-scale deformable image registration (DIR) framework with unsupervised joint training of convolutional neural network (MJ-CNN) for 4D-CT inter-phase registration. It was shown that, though being trained on a 4D-CT dataset, the MJ-CNN framework also performed well on both CT-CBCT and CBCT-CBCT registration without re-training or fine-tuning with a run time of about 1.4s for image size of 256 × 256 × 96.
	
	Recently, Deep learning (DL)-based medical image registration has become a hot research topic and achieved promising performances. Two thorough review papers on DL based image registration were recently published by Fu et al. \cite{RN16} and Haskins et al. \cite{RN17}. Generally, DL-based image registration methods can be divided into three categories: deep iterative registration, supervised transformation predication and unsupervised transformation prediction. The limitations of the deep iterative registration is that they inherit the iterative nature of conventional DIR methods \cite{RN18}, which slows the registration process. Supervised transformation prediction methods utilize the ‘ground truth DVF’ which is usually obtained using other DIR methods or artificially generated and thereafter quality checked by experts to supervise the network training \cite{RN19}. The quality control of the fidelity of the ‘ground truth DVF’ are subjective which may induce further inter-observer variability of the network performance. Unsupervised transformation prediction enables large number of datasets to be used in training since no ‘ground truth DVF’ is needed. However, without ground truth transformations, it is difficult to define proper loss functions of the networks. A spatial transformer network (STN) \cite{RN20} was proposed to generate the deformed image which enables image similarity loss calculation during the training process. 
	In this study, a novel unsupervised deep learning framework for DIR of inter-fraction CBCT images is proposed. Several strengths in the network design are considered in the proposed workflow: I) This work is based on the STN which explicitly enables the loss function to be defined without any manually aligned or pre-registered image pairs \cite{RN20, RN21}. Loss functions of image similarity and DVF regularization are used in the training stage of the proposed workflow. II) GAN architectures have been incorporated in the STN to improve the realism of the predicted DVFs.  III) A multi-scale framework was adopted to capture the coarse-scale and fine-scale motion.

	\noindent 
	\section{Methods and Materials}
	\noindent 
	\subsection{The proposed workflow of Spatial Transformation-Based Network}
	
	The schematic flowchart of the proposed method is outlined in Fig. 1. The training and inference stages follow the same feed forward path through a STN, which consists of a global generative adversarial network (GlobalGAN) and a local GAN (LocalGAN). The GlobalGAN is trained with the whole volume of the moving and the target images to capture the global geometric deformation and to generate a global deforming vector field (DVF) which facilitates the coarse alignment. The global DVF and moving CBCT fraction are then fed into spatial transformation to generate the globally deformed fractional CBCT images. However, the global DVF may fail to provide accurate local image registration due to non-rigid geometry and anatomic movements. To improve the accuracy of local registration, a LocalGAN is designed to capture the local deformation on top of the globally deformed CBCT images to match the target image. In the training of the LocalGAN, three-dimensional (3D) image patches with a size of 64 × 64 × 64 are extracted from the globally deformed CBCT images and the target CBCT images with an overlap size of 32 × 32 × 48. An image similarity loss and a regularization loss, as well as an adversarial loss, are included in the loss function of the GlobalGAN and LocalGAN. During the training stage, the GlobalGAN and LocalGAN are trained without the supervision of ground truth DVFs.
	
	\begin{figure}
		\centering
		\noindent \includegraphics*[width=6.50in, height=4.20in, keepaspectratio=true]{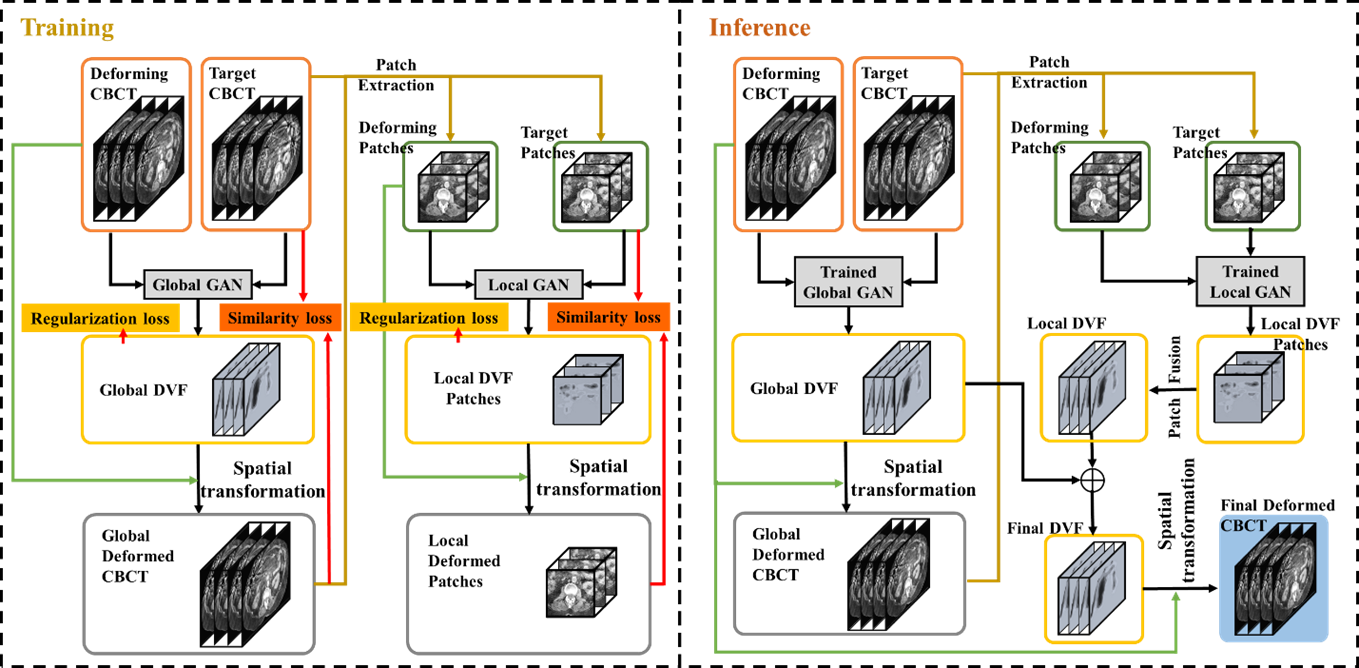}
		
		\noindent Figure 1. Schematic chart of the proposed workflow of unsupervised deep learning for inter-fraction CBCT deformable image registration (left column: training stage; right column: inference stage).
	\end{figure}

	\noindent 
	\subsubsection{Architectures of the Networks}
	
	The GlobalGAN and LocalGAN share similar GAN architectures with different learnable parameters. The structures of the generator and discriminator are shown in Fig. 2. In the generator network, image sizes of the input image pairs are reduced while being encoded through 11 convolutional layers. In order to up-sample the generated DVFs to matrix sizes the same as the input images, bilinear interpolation was applied. The discriminators in the two GANs are implemented as conventional fully convolution networks (FCN) \cite{RN22} for the necessary regularization to generate realistic DVFs. Since the discriminators are trained to distinguish the deformed images from the real CBCT images, they encourage the GlobalGAN and LocalGAN to predict realistic DVFs by penalizing unrealistic deformed images.
	
	\begin{figure}
		\centering
		\noindent \includegraphics*[width=6.50in, height=4.20in, keepaspectratio=true]{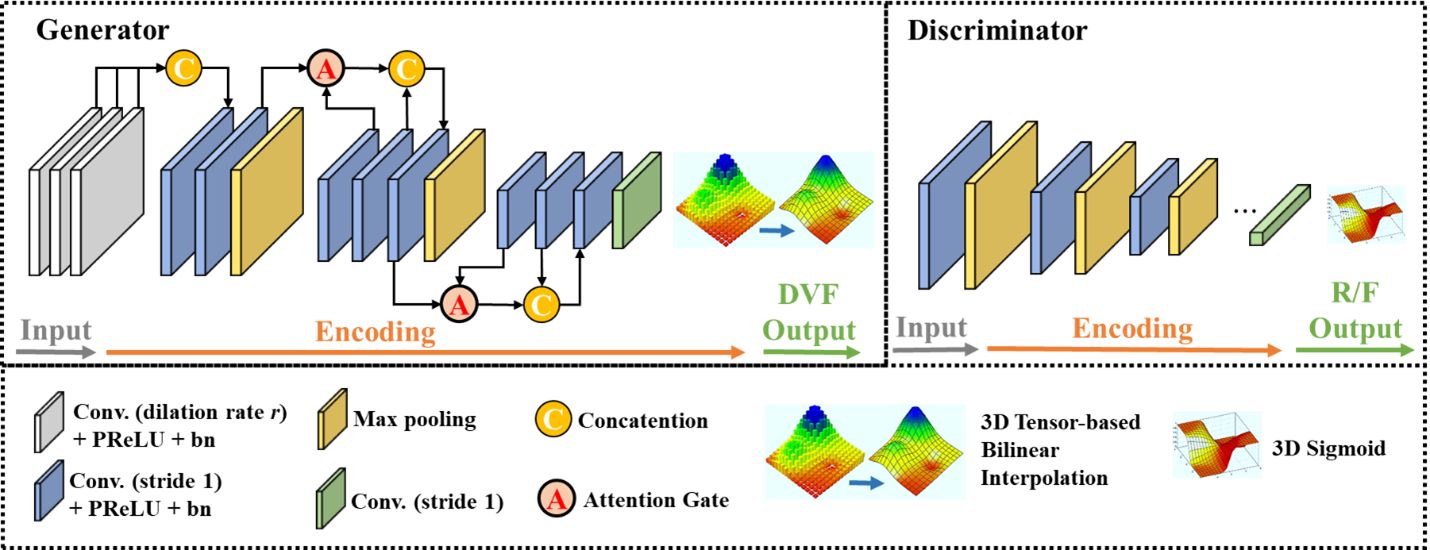}
		
		\noindent Figure 2. Network architecture of the generator (upper left) and discriminator (upper right) used in the GAN. The information of the layers is shown in the lower panel.
	\end{figure}
	
	\noindent 
	\subsubsection{Attention Gates}
	In order to force the model to focus on learning the motion information, self-attention network is integrated into GAN architectures for both GlobalGAN and LocalGAN. Self-attention network is constructed by integrating two attention gates into the generator between the convolution layers before and after the max pooling layers. With the attention gates, feature maps with different scales of the adjacent convolution layers are combined and operated right before concatenation. Attention gates have previously been explored in the context of semantic segmentation and were able to capture the most relevant semantic contextual information without using a very large reception field \cite{RN24, RN23}.
	
	\noindent 
	\subsubsection{Loss Functions}
	
	The image similarity, regularization and adversarial losses are consisted in the loss function of the GlobalGAN and LocalGAN. The difference between the loss term used for GlobalGAN and LocalGAN is that the loss of GlobalGAN is calculated based on whole images and the loss of LocalGAN is calculated based on patches.
	
	\begin{equation} 
	G=\underset{G}{arg\min}\left\{\alpha\cdot(SIM(I_d,I_t))+\beta\cdot ADV(I_d,I_t)+\gamma\cdot R(DVF)\right\}  
	\end{equation}
	
	where $DVF=G(I_m,I_t)$ represents the predicted DVF from a pair of moving $I_m$ and target $I_t$ images. The deformed image, $I_d=STN(I_m,DVF)$,  is obtained by warping the moving image patch by the predicted DVF \cite{RN25}. $SIM(I_d,I_t)$ denotes the image similarity loss.
	\begin{equation}
	SIM(I_d,I_t )=[1-NCC(MIND(I_d ),MIND(I_t ))]+\delta\cdot GD(MIND(I_d ),MIND(I_t ))
	\end{equation}
	where NCC($\cdot$) and GD($\cdot$) denote the normalized cross-correlation (NCC) loss and the gradient difference (GD) loss, respectively, between the deformed and target images. MIND($\cdot$) denotes the modality independent neighbourhood descriptor (MIND) \cite{RN26}. Due to scattering and image artefacts, inter-fraction CBCT HU values are inconsistent, which may deteriorate the effectiveness of similarity metrics such as mean square error and NCC. The MIND descriptor is a modality independent neighbourhood descriptor with normalized image intensity which is used here to pre-process the images before similarity measurement using NCC and GD.
	
	ADV($\cdot$) denotes the adversarial loss that is computed as the discriminator binary cross entropy loss of the deformed and target images. The purpose of the adversarial loss is to encourage the deformed image to approach realistic CBCT image by penalizing unreasonable DVFs and unrealistic deformed images.
	
	R(DVF) denotes the regularization term.
	\begin{equation}
		R(DVF)=\mu_1||\nabla DVF||_2+\mu_2 ||\nabla^2 DVF||_2
	\end{equation}
	Weighted ﬁrst and second derivatives of the DVF are included in the regularization term to enforce general smoothness of the predicted DVF. Values of $\mu_1$ and $\mu_2$ are set as 1 and 0.5 in this study, respectively. 
	
	The hyperparameters of $\alpha$,$\beta$,$\gamma$ and $\delta$ are empirically set as 200, 1, 10 and 5, respectively, according to numerical experiments.

	\noindent 
	\subsection{Datasets and Experiments}
	
	100 fractional CBCT images of 20 abdominal cancer patients who underwent radiotherapy were retrospectively investigated. The CBCT images have a resolution of 0.90 mm × 0.90 mm × 2.0 mm with size of 512 × 512 × 88. These images were acquired before each fraction during a five-fraction treatment course of each patient. Fiducial markers were implanted in the patients for tumor localization and external beam treatment planning. 
	
	The overall performance of the proposed method was investigated via a five-fold cross-validation. Specifically, the CBCT image data of the 20 patients was first randomly and equally separated into five groups, of which four groups were used for training; and the rest group was used for testing. The training and testing experiments were repeated five times by rotating each group as the testing group.
	
	In a holdout test, 105 fractional CBCTs from a cohort of 21 different abdominal cancer patients were investigated to evaluate the proposed method. The trained STN was tested on the holdout dataset without re-training or parameter fine-tuning.
		
	\noindent 
	
	\noindent 
	\subsection{Evaluations}
	
	Qualitative evaluations of the proposed method were performed by visually assessing the alignment between the target and deformed CBCT images. Both the fusion images and the absolute difference images between the target and deformed images were generated for the visual assessment. The absolute intensity difference profiles along a line in the anterior-posterior direction are also plotted to demonstrate the accuracy of image alignment. To demonstrate the efficacy of the integrated attention gates, DIR results with and without the attention gates were compared.
	
	For quantitative evaluations, mean absolute errors (MAEs) and normalized cross correlations (NCCs) between the target and the deformed CBCT images were calculated. The target registration errors (TREs) and dice similarity coefficients (DSCs) were also calculated. 
	
	The TRE was calculated as Euclidean distance between the landmark positions in the target and deformed CBCT images. The implanted fiducial markers with two or more additional landmarks identified by an experienced medical physicist are used for the TRE calculation. Four examples of the selected landmarks are shown in Fig. 3(a1) as red ‘×’ marks, of which two were implanted fiducials and the other two were the tips of the spinous process and rib bone. These landmarks were selected due to the fact that physicians usually prescribe to “match to target” or “match to bone” for on-treatment patient setup. With the position of the i-th landmark in patient K being denoted as $P_K^i$ in the moving fraction and the position of its corresponding landmark in the same patient denoted as $\hat{P}_K^i$, the TRE was calculated as:
	\begin{equation}
		TRE(i,K)=||P_K^i-\hat{P}_K^i||_2
	\end{equation}
	where $||\cdot||_n$ stands for the L-n norm of the matrix.
	
	The MAE for patient K was calculated as 
	\begin{equation}
		MAE(K)=\frac{1}{||B_K||_0}||B_K (I_{K,d}-I_{K,t})||_1
	\end{equation}
	where the target and deformed images of patient K are denoted as $I_{K,t}$ and $I_{K,d}$. $B_K$ stands for the image mask of the bounding box of the patient body, which was determined by all the tissues/organs with HU values higher than -300 HU in this study. 
	
	The NCC for patient K was calculated as
	\begin{equation}
		NCC(K)=\frac{\Sigma_{x,y,z} [B_K (I_{K,d} )-\overline{B_K (I_{K,d} ) }]\cdot[B_K (I_{K,t} )-\overline{B_K (I_{K,t} ) }]}{\left\{\Sigma_{x,y,z} [B_K (I_{K,d} )-\overline{B_K (I_{K,d} )}]^2 \Sigma_{x,y,z}[B_K (I_{K,t} )-\overline{B_K (I_{K,t})}]^2 \right\}^{1/2}}
	\end{equation}
	where $\Sigma_{x,y,z}$ stands for elemental wise summation of the 3D CBCT images; and $\overline{\cdot}$stands for the mean of the image. $B_K$ has the same meaning as in the calculation of MAE of Eq. 5.

	The DSC for patient K was calculated as 
	\begin{equation}
	DSC(K)=\frac{2||M_{K,d}\cdot M_{K,t} ||_0}{||M_{K,d} ||_0+||M_{K,t} ||_0 }
	\end{equation}
	where $M_{K,d}$ and $M_{K,t}$ are the binary mask of the bony structures in the deformed and target images. $M_{K,d}\cdot M_{K,t}$ stands for the element-wise multiplication of these masks. The masks of the bony structures for the DSC calculation in this study were determined by the tissues/organs with HU values higher than 300 HU.

	\noindent 
	
	\section{Results}
	
	Fractional CBCT images of an abdominal cancer patient during the treatment course are shown in Fig. 3 to demonstrate the overall registration results of the proposed method. In Fig. 3, the first fraction is shown in the upper left corner as the target CBCT image; and the subsequent four fractions (the following four columns) that were registered with on-treatment manual rigid registration, the proposed method without attention gates in the generators of the GlobalGAN and LocalGAN and the proposed method are shown in the first to third rows, respectively. Three observations could be drawn from Fig. 3: 1) The stomach region appearances could be essentially different in each treatment fraction (pointed by orange arrows); 2) The overall body (anterior-posterior) location could be well resemble to the target image with the proposed method, but not the manual rigid registration or the proposed method without attention gates (regions pointed by yellow arrows); 3) The anatomical morphology of the internal organs, fiducial markers, and even the streak artifacts, are also changed/modified, which is believed to be induced by the LocalGAN. Overall, the proposed method has better global and local alignment due to the multi-scale registration scheme realized by the GlobalGAN and Local GAN.
	
	\begin{figure}

		\noindent \includegraphics*[width=6.50in, height=4.20in, keepaspectratio=true]{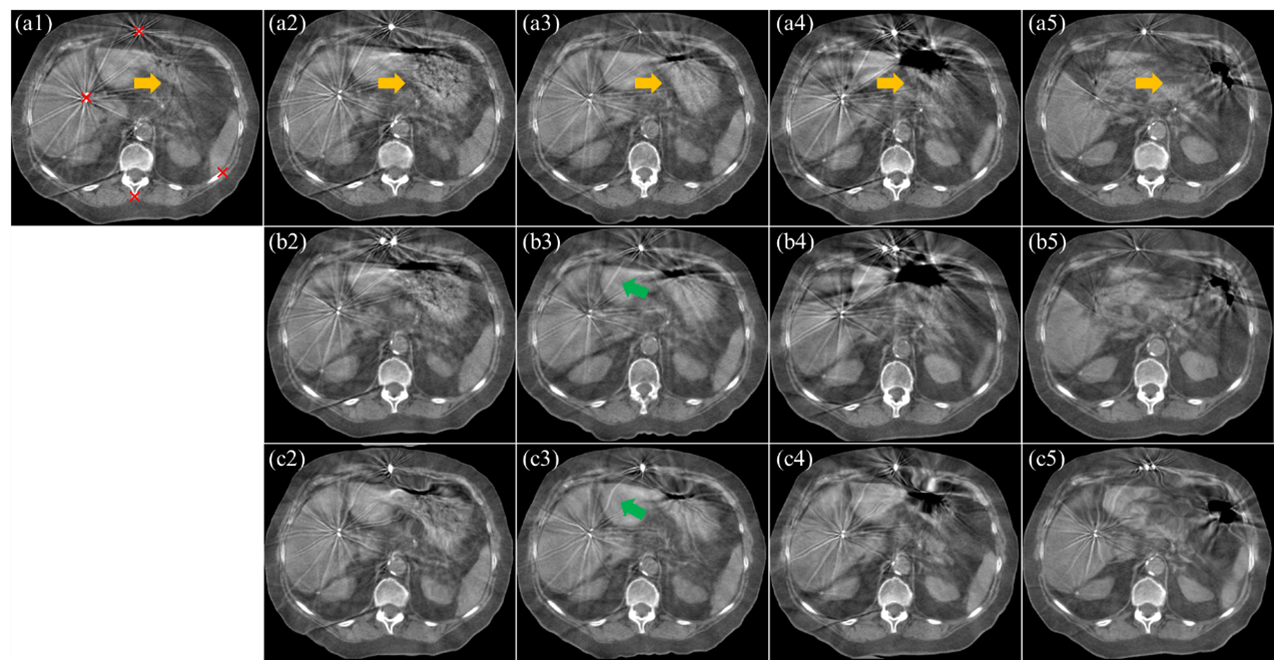}
		
		\noindent Figure 3. CBCT images of a cancer patient acquired in a five-fraction treatment course. The first fraction is treated as the target and shown in subfigure (a1). The rest four fractions (second to fifth) are treated as the moving images and shown in the second to fifth columns. The on-treatment manual rigid registered images of these fractions are shown in the first row (a2 – a5). The registered images with the proposed method without the attention gates in the generators are shown in the second row (b2 – b5). The registered images of the second to the fifth fractions with the proposed method are shown in the third row (c2 – c5). Four landmarks for target registration error (TRE) calculation are shown as red cross marks in (a1). Green arrows indicate streak artifacts that were warped by the deformable registration algorithms. Orange arrows indicate large differences between inter-fraction CBCT images. The image showing window is [-300 300] HU.
	\end{figure}

	\begin{figure}

		\noindent \includegraphics*[width=6.50in, height=4.20in, keepaspectratio=true]{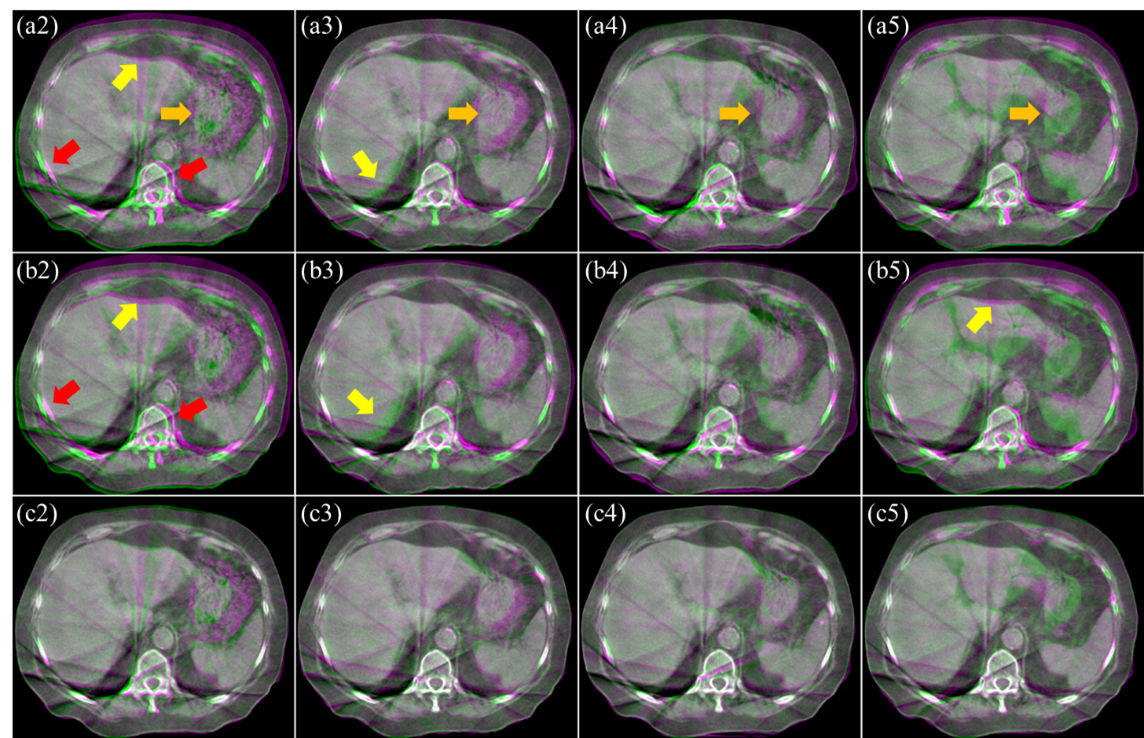}
		
		\noindent Figure. 4. Fusion images of the target fraction and the registered fractions of the patient shown in Fig. 3 at a different slice location. The fusion images of first (target) fraction and the second to fifth fractions with on-treatment manual rigid registration are shown in the first row (a2 – a5), respectively. The fusion images of the target fraction and those registered with the proposed method without attention gates are shown in the second row (b2 – b5). The fusion images of the target fraction and those registered with the proposed method are accordingly shown in the third row (c2 – c5).
	\end{figure}

	For better visibility of the registration quality, the fusion images of the registered fractions with the target fraction of the same patient shown in Fig. 3 at a different slice location are shown in Fig. 4. In the fusion images, the deformed images and the target image are show in the red and green channels, respectively; and they are labeled as the same subfigures as in Fig. 3 for consistence. Since the stomach region appearance was different from fraction to fraction, suboptimal image registration quality can be found (pointed by orange arrows). As shown by the yellow arrows in Fig. 4, with the proposed method, the liver area could be well registered to the target region, but not with the other two methods. Bony structures are also usually treated as indicators of image registration quality. For the ribs and the spine, (regions pointed by red arrows), the proposed method (c2 – c5) outperformed the other two approaches (a2 – a5, b2 – b5). 
	
	The difference images between the registered images (a2 – a5, b2 – b5 and c2 – c5) and the target images (a1) in Fig. 4 are shown in Fig. 5. The difference images are labeled as the same subfigures as in Fig. 4; and they have the same arrows copied from Fig. 4. The proposed method has shown better alignment both globally and locally, evidenced by the lower HU difference, than the other two methods. Despite of the challenges at the stomach region, the proposed method shows the lowest HU difference (orange arrows). Consistent with previous observations, the proposed method outperformed the other two methods at the liver and spine boundaries, as indicated by the yellow and red arrows.

	\begin{figure}
		\centering
		\noindent \includegraphics*[width=6.50in, height=4.20in, keepaspectratio=true]{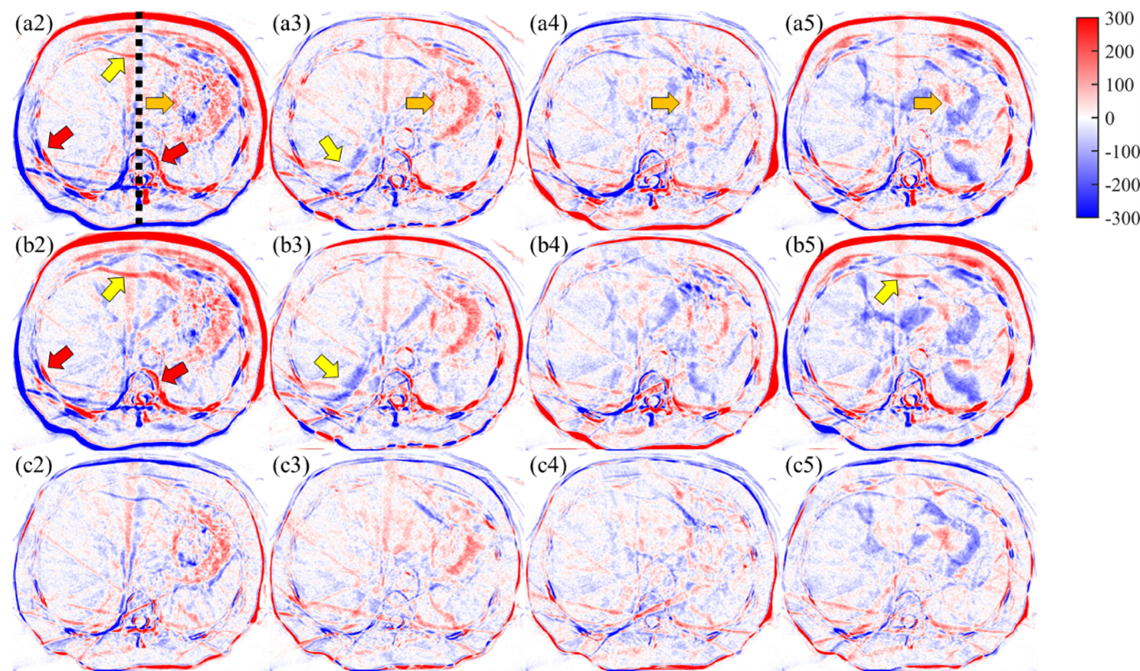}
		
		\noindent Figure. 5. Difference images of the registered second to fifth fractions versus the target fraction of the patient shown in Fig. 4. The difference images of the second to fifth fractions with on-treatment manual rigid registration are shown in the first row (a2 – a5), respectively. The difference images of the registered fractional images with the proposed method without attention gates are shown in the second row (b2 – b5). The difference images of those registered with the proposed method are accordingly shown in the third row (c2 – c5).
	\end{figure}

	\begin{figure}
	
	\noindent \includegraphics*[width=6.50in, height=4.20in, keepaspectratio=true]{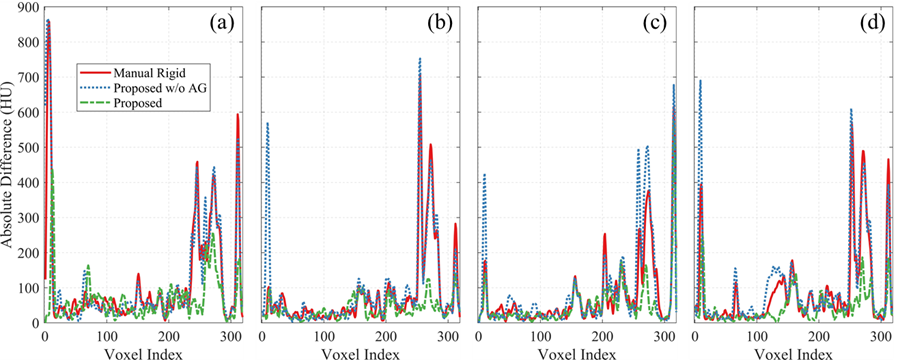}
	
	\noindent Figure. 6. Absolute HU difference profile along the black dotted line in Fig. 5. The registered images of the second to the fifth fractions versus that of the first fraction are shown in subfigures (a) – (d) respectively. In each subfigure, the absolute HU difference profile of the manual rigid registration is plotted as solid line; profiles of the proposed methods with and without attention gates were plotted as dash-dotted and dotted lines, respectively.
	\end{figure}

	Shown in Fig. 6 are the absolute HU difference profiles along the black dotted line in Fig. 5. The large HU difference anteriorly (voxel index < 20) and posteriorly (voxel index > 300) were caused by the body contour misalignment; and those near the spine (240 < voxel index < 300) were caused by the spine misalignment Compared with the other two methods, the proposed method has the best performance with lowest HU differences. 
	
	The average TREs over all patients are listed in Table 1. The average TREs is less than 2 mm for the proposed method, much better than the other two methods. The smaller TRE standard deviations of the proposed method also demonstrate better registration robustness. Two sample t-tests on the TRE results of the three investigated methods showed that the TRE improvement of the proposed method over the other two is statistically significant. On the contrary, the TRE improvement of the proposed method without attention gates over the on-treatment manual rigid registration was not statistically significant, showing the efficacy of the attention gates.
	
	The average MAEs between the deformed CBCT images and the target images are listed in Table 2. Same conclusion as the TRE analysis can be drawn, which strengthened our argument that the proposed method with attention gates has the best performance out of the three methods.
	
	As two important metrics indicating the image registration quality, the NCCs and DSCs were calculated and shown in Tables 3 and 4, respectively. From these two tables, same conclusion that the proposed method substantially outperformed the other two methods with statistical significance can be drawn.
	
	Table 5 shows the average TREs, MAEs, NCCs and DSCs for all the patients in the holdout dataset. The uniform improvement of the proposed method over the other two methods across all four metrics demonstrated that the proposed method has great robustness and generalizability on the holdout datasets.

\begin{table}[htbp]
	\centering
	\caption{Target registration errors (TREs) of the registered second to fifth fractions, as well as overall TRE regardless of fraction, versus the target fraction. The TREs are calculated over all of the involved patients in the experiment. (Unit: mm)}
	\begin{tabular}{@{}lllllll@{}}
		\toprule
		Fraction & Manual Rigid 1 & Proposed w/o AG 2 & Proposed 3 & \begin{tabular}[c]{@{}l@{}}p value\\    (1 vs 2)\end{tabular} & \begin{tabular}[c]{@{}l@{}}p value\\    (1 vs 3)\end{tabular} & \begin{tabular}[c]{@{}l@{}}p value\\    (2 vs 3)\end{tabular} \\ \midrule
		2        & 4.85±2.73      & 4.46±2.34         & 1.92±0.97  & 0.612                                                         & \textless{}0.001                                              & \textless{}0.001                                              \\
		3        & 4.24±2.69      & 3.66±2.26         & 1.74±0.95  & 0.409                                                         & \textless{}0.001                                              & \textless{}0.001                                              \\
		4        & 6.28±3.72      & 5.35±3.66         & 1.99±1.14  & 0.379                                                         & \textless{}0.001                                              & \textless{}0.001                                              \\
		5        & 4.88±2.70      & 5.07±3.35         & 1.97±1.59  & 0.836                                                         & \textless{}0.001                                              & \textless{}0.001                                              \\
		Overall  & 5.07±3.05      & 4.64±3.01         & 1.91±1.18  & 0.320                                                         & \textless{}0.001                                              & \textless{}0.001                                              \\ \bottomrule
	\end{tabular}
\end{table}%

\begin{table}[htbp]
	\centering
	\caption{Mean absolute errors (MAEs) of the registered second to fifth fraction images, as well as overall MAE regardless of fraction, versus the target fraction image. The MAEs are calculated over all of the involved patients in the experiment. (Unit: HU)}
	\begin{tabular}{@{}lllllll@{}}
		\toprule
		Fraction & Manual Rigid 1 & Proposed w/o AG 2 & Proposed 3 & \begin{tabular}[c]{@{}l@{}}p value\\    (1 vs 2)\end{tabular} & \begin{tabular}[c]{@{}l@{}}p value\\    (1 vs 3)\end{tabular} & \begin{tabular}[c]{@{}l@{}}p value\\    (2 vs 3)\end{tabular} \\ \midrule
		2        & 48.25±13.80    & 44.42±10.04       & 34.44±7.73 & 0.359                                                         & 0.013                                                         & 0.007                                                         \\
		3        & 45.29±10.74    & 41.40±10.07       & 33.35±8.63 & 0.331                                                         & 0.016                                                         & 0.032                                                         \\
		4        & 44.45±9.83     & 42.74±9.07        & 33.09±7.49 & 0.620                                                         & 0.009                                                         & 0.005                                                         \\
		5        & 43.15±9.40     & 42.64±8.95        & 32.81±7.08 & 0.881                                                         & 0.012                                                         & 0.004                                                         \\
		Overall  & 45.28±10.81    & 42.80±9.47        & 33.42±7.48 & 0.179                                                         & \textless{}0.001                                              & \textless{}0.001                                              \\ \bottomrule
	\end{tabular}
\end{table}%

\begin{table}[htbp]
	\centering
	\caption{Normalized cross correlations (NCCs) of the registered second to fifth fractions, as well as overall NCC regardless of fraction, versus the target fraction. The NCCs are calculated over all of the involved patients in the experiment. Only tissues/organs with voxel intensity higher than -300 HU are considered.}
	\begin{tabular}{@{}lllllll@{}}
		\toprule
		Fraction & Manual Rigid 1 & Proposed w/o AG 2 & Proposed 3 & \begin{tabular}[c]{@{}l@{}}p value\\    (1 vs 2)\end{tabular} & \begin{tabular}[c]{@{}l@{}}p value\\    (1 vs 3)\end{tabular} & \begin{tabular}[c]{@{}l@{}}p value\\    (2 vs 3)\end{tabular} \\ \midrule
		2        & 0.88±0.06      & 0.89±0.05         & 0.93±0.04  & 0.345                                                         & \textless{}0.001                                              & \textless{}0.001                                              \\
		3        & 0.90±0.06      & 0.90±0.06         & 0.94±0.04  & 0.815                                                         & \textless{}0.001                                              & \textless{}0.001                                              \\
		4        & 0.89±0.04      & 0.89±0.05         & 0.94±0.03  & 0.963                                                         & \textless{}0.001                                              & \textless{}0.001                                              \\
		5        & 0.90±0.05      & 0.89±0.04         & 0.94±0.04  & 0.230                                                         & \textless{}0.001                                              & \textless{}0.001                                              \\
		Overall  & 0.89±0.05      & 0.89±0.05         & 0.94±0.04  & 0.880                                                         & \textless{}0.001                                              & \textless{}0.001                                              \\ \bottomrule
	\end{tabular}
\end{table}%

\begin{table}[htbp]
	\centering
	\caption{Dice similarity coefficients (DSCs) of the registered second to fifth fractions, as well as overall DSC regardless of fraction, versus the target fraction. The DSCs are calculated over all of the involved patients in the experiment. Only bony tissues/organs with voxel intensity higher than 300 HU are considered.}
	\begin{tabular}{@{}lllllll@{}}
		\toprule
		Fraction & Manual Rigid 1 & Proposed w/o AG 2 & Proposed 3 & \begin{tabular}[c]{@{}l@{}}p value\\    (1 vs 2)\end{tabular} & \begin{tabular}[c]{@{}l@{}}p value\\    (1 vs 3)\end{tabular} & \begin{tabular}[c]{@{}l@{}}p value\\    (2 vs 3)\end{tabular} \\ \midrule
		2        & 0.34±0.12      & 0.36±0.10         & 0.51±0.10  & 0.536                                                         & 0.004                                                         & \textless{}0.001                                              \\
		3        & 0.36±0.12      & 0.37±0.11         & 0.52±0.11  & 0.716                                                         & 0.005                                                         & \textless{}0.001                                              \\
		4        & 0.35±0.08      & 0.36±0.08         & 0.52±0.07  & 0.872                                                         & \textless{}0.001                                              & \textless{}0.001                                              \\
		5        & 0.36±0.11      & 0.37±0.10         & 0.52±0.09  & 0.878                                                         & 0.003                                                         & \textless{}0.001                                              \\
		Overall  & 0.35±0.10      & 0.37±0.10         & 0.52±0.09  & 0.495                                                         & \textless{}0.001                                              & \textless{}0.001                                              \\ \bottomrule
	\end{tabular}
\end{table}%

\begin{table}[htbp]
	\centering
	\caption{The overall TREs, MAEs, NCCs and DSCs of the registered CBCTs, regardless of fraction, versus the target CBCTs among all the patients in the holdout dataset. The methods for metric calculations are consistent with those in Tables 1 – 5.}
	\begin{tabular}{@{}lllllll@{}}
		\toprule
		\begin{tabular}[c]{@{}l@{}}Overall\\    Metrics\end{tabular} & Manual Rigid 1 & Proposed w/o AG 2 & Proposed 3 & \begin{tabular}[c]{@{}l@{}}p value\\    (1 vs 2)\end{tabular} & \begin{tabular}[c]{@{}l@{}}p value\\    (1 vs 3)\end{tabular} & \begin{tabular}[c]{@{}l@{}}p value\\    (2 vs 3)\end{tabular} \\ \midrule
		TRE (mm)                                                     & 5.12±2.82      & 4.96±2.40         & 2.34±1.74  & 0.229                                                         & \textless{}0.001                                              & \textless{}0.001                                              \\
		MAE   (HU)                                                   & 50.74±9.34     & 49.34±9.15        & 38.83±7.88 & 0.499                                                         & \textless{}0.001                                              & 0.001                                                         \\
		NCC                                                          & 0.89±0.04      & 0.90±0.04         & 0.92±0.02  & 0.037                                                         & \textless{}0.001                                              & \textless{}0.001                                              \\
		DSC                                                          & 0.41±0.10      & 0.43±0.09         & 0.51±0.04  & 0.341                                                         & \textless{}0.001                                              & \textless{}0.001                                              \\ \bottomrule
	\end{tabular}
\end{table}%

	\bigbreak
	
	\noindent 
	\section{Discussion}
	
	Over the course of radiotherapy treatment, a method to provide fast and accurate inter-fraction CBCT image registration is essential for evaluation of the geometric and anatomic changes. With the proposed registration tool, quantitative anatomic changes could be calculated for inter-fractional variation modeling and prediction. It could potentially inform the physician in future treatment planning such as targets margin definition and image guidance usage frequency, tradeoff between target coverage and OAR sparing. The proposed inter-fraction CBCT image registration could also enable many applications such as image segmentation \cite{RN27}, motion estimation \cite{RN7, 28}, image fusion \cite{RN29} and treatment response evaluations \cite{RN30, 31}. Deep learning-based DIR is promising for the online DIR task of large volume CBCT images in radiotherapy. In this work, an unsupervised deep learning based inter-fraction CBCT registration method, which takes less than 3 seconds to perform a CBCT-CBCT registration, is proposed and its feasibility and performance are investigated through qualitative and quantitative evaluations. The proposed method can also perform well on the holdout dataset without re-training. The major contributions of the proposed workflow can be summarized as:
	
	a)	A multi-scale unsupervised deep learning-based DIR method is proposed for inter-fraction CBCT DIR in image-guided radiotherapy. The unsupervised training of the proposed STN-based network overcomes the challenge of collecting large amount of ground truth datasets via either manually aligning, which is labor-intense, or artificially synthesizing, which is error prone. The integration of GlobalGAN and LocalGAN networks captures the image misalignment in a multi-scale manner; and its effects for the DIR can be observed in Figs. 3 – 6. 
	
	b)	The self-attention network enabled by the attention gates in the generator is essential to learn the differences between major and minor motion regions and avoid deforming the bony structures. The attention gate operation helps to retain only relevant activations and remove irrelevant/noisy responses. The neuron activations during both the forward and backword passes are filtered by the attention gates. In addition, gradients originating from image background during the backward pass are down weighted. The self-attention network allows model parameters in shallow layers to be updated based on spatial regions that are most relevant to a given task, i.e. motion estimation, such that the generator network could gain the ability to highlight the features from previous layers and to well represent the image motion. Thus, the integration of attention gates in the generator network could essentially improve the capturing of structural differences between the moving and target images.
	
	c)	Adversarial network is integrated into the proposed framework to enforce additional DVF regularization by penalizing unrealistic deformed images. Since the networks are designed to be trained in an unsupervised manner, DVF regularization is necessary to generate realistic results. Smoothness constraint has been commonly used in the literature for DVF regularization \cite{RN7}. However, the smoothness constraint alone is insufficient for realistic DVF prediction especially when the network is trained in a completely unsupervised manner. Therefore, for additional DVF regularization, a discriminator is proposed to be integrated into STN. Since the purpose of the discriminator is to better differentiate deformed images from target images, as such unrealistic deformed images are penalized. Realistic DVFs are then encouraged to be predicted by the generator. The speed of the inference stage will not be affected as the discriminator is only used in the training stage. 

	To generate DVFs with the same matrix sizes as the input images, since image sizes of the input image pairs are reduced while being encoded through 11 convolutional layers in the generator network, bilinear interpolation was used to up-sample the DVFs. As an alternative, transpose-convolution layers with trainable parameters can be used to up-sample the DVFs \cite{RN32}. However, we have found that bilinear interpolation, which does not contain trainable parameters, performs much better than the transpose-convolution layers in predicting accurate DVFs. The reason of this might be that bilinear interpolation tends to generate smooth DVFs which are desired in medical image registration. On the other hand, the transpose-convolution layers often generate unrealistic DVFs even with heavily-weighted DVF smoothness regularization term \cite{RN21, 33}. 
	
	No prepossessing had been applied on the fractional CBCT images to improve the image quality before deformable image registration using the proposed method. Therefore, suboptimal image quality of the CBCT images, such as inter-fraction variations in the altered HU values and streak artifacts, could impact the accuracy of image registration. Deep learning-based image synthetic approaches have been investigated and shown promising result in improving the image quality \cite{RN34}. Then, incorporating of these approaches with the proposed method might improve the image registration result. Further investigations on this topic are needed in future works. 
	
	It was observed that the inter-fraction shape and position of the organs in some abdominal patients may vary significantly due to gas fillings, bowel movements and/or respiratory motions. This is an extremely challenging situation for both unsupervised and supervised deep learning image registration, because that significant shape and position variations may become beyond model ability of the proposed method and may also impossible to manual labelling of the ground truth. The proposed method may fail to accurately register the inter-fraction CBCTs of these patients. Future researches are necessary for this situation.
	
	In this study, only a few qualitative (visual inspections on the deformed, fusion, difference images and intensity profiles) and quantitative (TRE, MAE, NCC and DSC) evaluations were performed. These metrics are not directly related with clinical outcomes. It is anticipated that more clinical investigations, i.e. dose volume histograms, patient follow-ups, etc., are needed in order to concluded whether the proposed method could be applied and effective in the image-guided radiotherapy.

	\bigbreak
	
	\noindent 
	\section{Conclusion}
	
	An unsupervised deep learning-based CBCT-CBCT registration method is developed and its feasibility and performance are investigated. The proposed method is able to accurately register images between the moving and target CBCT fractions within three seconds in a single forward network prediction, as such it is expected to be promising as a fast and straightforward image registration tool for motion management and treatment planning in image-guided radiotherapy.

	\noindent 
	\bigbreak
	{\bf ACKNOWLEDGEMENT}
	
This research is supported in part by the National Institutes of Health under Award Number R01CA215718, and R01EB028324.

	\noindent 
	\bigbreak
	{\bf Disclosures}
	
	The authors declare no conflicts of interest.

	\noindent 
	
	\bibliographystyle{plainnat}  
	\bibliography{arxiv}      
	
\end{document}